\documentclass[checkin,showpacs,aps,pra,nofootinbib]{revtex4-1}
\usepackage{amsmath}
\usepackage{graphicx,color}
\usepackage{pifont}
\usepackage{amssymb}
\usepackage{datetime}
\usepackage{setspace}

\newcommand{\nab}{\nabla}

\newcommand{\clr}{}

\newcommand{\breaker}{\nonumber \\ }

\newcommand{\bn}{\begin{enumerate}}
\newcommand{\en}{\end{enumerate}}
\newcommand{\cE}{\boldsymbol{\cal E}}

\newcommand{\bG}{{\bf G}}
\newcommand{\bq}{{\bf q}}
\newcommand{\bQ}{{\bf Q}}
\newcommand{\br}{{\bf r}}
\newcommand{\bR}{{\bf R}}

\newcommand{\hH}{\hat{H}}

\newcommand{\hT}{\hat{T}}

\newcommand{\hc}{\hat{c}}

\newcommand{\hcp}{\hat{c}^\dag}

\newcommand{\bj}{{\bf J}}

\newcommand{\hp}{\hat{\bf p}}

\newcommand{\bk}{{\bf k}}
\newcommand{\bkr}{{\bf k,\br}}
\newcommand{\bqr}{{\bf q,\br}}

\newcommand{\hh}{\hat{H}}

\newcommand{\ba}{\begin{eqnarray}}
\newcommand{\ea}{\end{eqnarray}}

\usepackage{graphicx}

\renewcommand{\k}{\bf k}

\newcommand{\be}{\begin{equation}}
\newcommand{\ee}{\end{equation}}
\newcommand{\la}{\langle}

\newcommand{\ra}{\rangle}

\newcommand{\hr}{\hat{\rho}}

\begin{document}



\title{Generalized length-gauge time-dependent Liouville equation}


\author{G. P. Zhang$^*$} \affiliation{Department of Physics, Indiana
  State University, Terre Haute, IN 47809} \date{\today}

\author{M. S. Si$^{\dagger}$}
 \affiliation{School of Materials and Energy,
  Lanzhou University, Lanzhou 730000, China}

\begin{abstract}
  {
   The length-gauge (LG) time-dependent Liouville equation is a major
   workhorse to simulate ultrafast dynamics. The equation features the
   commutator between the density operator $\hr$ and the position
   operator $\br$, $\la n\bk|[\br,\hr]|m\bk'\ra$, where $|n\bk\ra$ and
   $|m\bk'\ra$ are Bloch states for bands $n$ and $m$ at crystal
   momenta $\bk$ and $\bk'$, respectively.  This commutator has two
   irreconcilable versions: Kohn and Luttinger (KL) (Phys. Rev. {\bf
     108}, 590 (1954)) have two partial derivatives with respect to
   $\bk$ and $\bk'$, but Aversa and Sipe (AS) (Phys. Rev. B {\bf 52},
   14636 (1995)) have one partial derivative and one Dirac delta.  For
   several decades, this outstanding discrepancy has received little
   attention.  Here, we show that the generalized LG time-dependent
   Liouville equation has two partial derivatives, but no Dirac delta,
   even when whether the electric field is homogeneous. Physically,
   the $\bk$-off-diagonal matrix elements of the position operator
   provide a channel by which the electric field generates
   $\bk$-off-diagonal matrix elements for $\hr$ in LG. The discrepancy
   emerges when the position matrix is assumed to be diagonal in
   $\bk$.  Consequently the AS' equation is a limiting case of our
   generalized Liouville equation, where the relative momentum $\bQ$
   between two states is ignored.  Our numerical test reveals that the
   $\bk$-diagonal formalism yields qualitatively different results and
   underestimates both energy absorption and polarization by a factor
   of 7.  Our study establishes a general formalism for the
   length-gauge time-dependent Liouville equation and opens the door
   for future research on the ultrafast time scale, while prior
   results based on the $\bk$-diagonal only equation may need to be
   revisited.  }
\end{abstract}

\maketitle

\section{Introduction}

\newcommand{\Ba}{ \begin{align} }


The traditional nonlinear optics and ultrafast dynamics in solids
often adopt two approaches \cite{shen,boyd}. One is based on the
time-dependent Schr\"odinger equation \cite{adams1956}.
 The other is based on the density
matrix \cite{boyd}, where one solves the time-dependent Liouville
equation, \be i\hbar \frac{\partial \hr}{\partial t}
=[\hh,\hr]=[\hh_0+\hh_I,\hr],
\label{eq1}
\ee where $\hr$ is the density operator, $\hh_0$ is the
nonperturbative Hamiltonian of the system and $\hh_I$ is the
interaction Hamiltonian between the light and the system.  In solids,
one often adopts the crystal momentum $\bk$ representation (CMR)
\cite{adams1959}, where the Bloch waves of an unperturbed crystal,
$\psi_n(\bk,\br)=u_{n\bk}(\br)e^{i\bk\cdot \br}$, are chosen as the
basis functions. Here $u_{n\bk}(\br)$ is the cell periodic part of the
wavefunction.  In the velocity gauge (VG) within the dipole
approximation, $\hh_I$ is the product of the vector potential and the
momentum \cite{jpcm23}, so it is diagonal in $\bk$, but $\bk$ is now
time-dependent \cite{kittel,krieger1986}.

In the length gauge (LG), one has \cite{adams1956} $\hh_I=e\br\cdot
\cE(t)$, where $e$ is the elementary charge, $\br$ is the position of
the electron, and $\cE(t)$ is the electric field of the light.
However, $\br$ is not translationally invariant, so it is customary to
convert it to a differential form, see more recent discussions
\cite{epl25}.  Aversa and Sipe (AS) \cite{aversa1995} adopted this
approach. They split the position operator into the intraband and
interband parts.  For convenience, we quote their equations 3 and 4
here (see further details in their paper), $\la
n\bk|\br_i|m\bk'\ra=\delta_{nm}[\delta(\bk-\bk')\xi_{nn}+i\nab_\bk
  \delta (\bk-\bk')],\la
n\bk|\br_e|m\bk'\ra=(1-\delta_{nm})\delta(\bk-\bk')\xi_{nm}, $ where
$\br_i$ refers to the intraband and $\br_e$ refers to the interband
parts, respectively. $\br_e$ is diagonal in $\bk$ due to the
periodicity of $u_{n\bk}(\br)$.  $\xi_{nm}(\bk)$ is the Berry
connection given as $
\xi_{nm}(\bk)=\frac{(2\pi)^3i}{\Omega}\int_\Omega d\br u^*_{n\bk}(\br)
\nab_\bk u_{m\bk}(\br), $  where $\Omega$ is the
volume.  Then they claimed that the matrix elements of the commutator
$[\br,S]$ between a simple operator $S$ and the position operator
$\br$ can be written as (their Eqs. 8 and 9) \be \la
n\bk|[\br_i,S]|m\bk'\ra \overset{?}{=}
i\delta(\bk-\bk')(S_{nm})_{;\bk}.
\label{aversa1}\ee
In their paper, they did not explain why this commutator yields a
Dirac $\delta(\bk-\bk')$, but gave the superoperator $(S_{nm})_{;\bk}$
as $(S_{nm})_{;\bk}= \frac{\partial S_{n,m}(\bk)}{\partial \bk}
-iS_{nm}(\bk)(\xi_{nn}(\bk)-\xi_{mm}(\bk)). $ As a result, their
Liouville equation (their equation 20) is now diagonal in $\bk$, which
has been extensively used in the literature {\clr
  \cite{pedersen2015,bonabi2017,parker2019,yue2020,liebscher2021,passos2021,thong2021,hannes2022,parks2023,hagen2023,sarkar2025,song2026,murotani2026,froese2026}.}

Historically, if the electric field is spatially inhomogeneous, the
density operator $\hr$ has off-diagonal matrix elements at two
different $\bk$'s \cite{hess1996,bufler1994}.  However, in LG $\hr$
intrinsically has both $\bk$-diagonal and $\bk$-off-diagonal elements.
Even for a homogeneous electric field, the position operator $\br$
still does not have translational symmetry, and must have off-diagonal
elements in $\bk$ \cite{epl25}.  For the last three decades AS'
commutator (Eq. \ref{aversa1}) has not been questioned
\cite{golde2008}, to the best of our knowledge.  When working on an
earlier study \cite{epl25}, we became aware of a similar commutator
between the density operator and the position operator.  Kohn and
Luttinger (KL) \cite{kohn1957} employed a {\it homogeneous} electric
field and showed that a similar commutator, $C=[\hr,\hh_1]$ (see the
first line below Eq. 21 in their paper), has matrix elements (their
Eq. 113) as \ba C_{ll'}=ieE_\alpha^{(0)}\left [\left (
  \frac{\partial}{\partial k_\alpha}+ \frac{\partial}{\partial
    k_\alpha'} \right ) \rho_{ll'}
           +\sum_{l''}(J_\alpha^{ll''}\rho_{l''l'}-\rho_{ll''}J_\alpha^{l''l'})
           \right ] \label{kl1} \ea which has two partial derivatives, one with respect
         to $k_\alpha$ and the other to $k'_\alpha$, but has no Dirac
         delta.
Here $\hh_1$ is just $\hh_I$
above.  KL used $x_\alpha$ in place of $\br$, with $\alpha$ denoting
the direction of $\br$.
          By contrast, Eq. \ref{aversa1} only
           has one partial derivative and one Dirac delta.  These
           disagreements are alarming, but did not seem to bother AS
           as they also cited KL's paper.  In Eq. \ref{kl1} and also
         in the KL paper, $l$ stands for the index pair $(\nu,\bk)$
         (see their paper for details).  $J_\alpha^{ll'}$ is given by
         \be J_\alpha^{ll'}=\frac{1}{\omega} \left ( \int_\omega
         w_{\nu \bk }^* \frac{\partial w_{\nu'\bk}}{\partial
           k_\alpha}d\br \right ) \delta_{\bk,\bk'},
\label{kohn1}
%
%
%
\ee which $\omega$ is the volume, same as $\Omega$ above, and $w_{\nu
  \bk}$ is the periodic part of the Bloch wavefunction, same as
$u_{n\bk}$ above.  $J_\alpha^{ll'}$ differs from the Berry
connection by $i$, though clearly Kohn and Luttinger were earlier to
find this relation.

One year later, Luttinger \cite{luttinger1958} provided another matrix
elements for the commutator between any operator $\Lambda$ and the
position operator $x_\alpha$ as (his equations B.20 and B.21)
$[\Lambda,x_\alpha]_{ll'}=-i[D_\alpha
  \Lambda_{ll'}+(J_\alpha,\Lambda)_{ll'}]$, where
$D_\alpha=\partial/\partial k_\alpha+\partial/\partial k'_\alpha$.
Note again that the commutator has two derivatives \cite{thomas1966}.
This important double-derivative has been absent from hundreds of
papers over the last half-century
\cite{blount1962,lei1984,krieger1986,aversa1995,iafrate2017}, even
absent from standard textbooks
\cite{kittel3,callaway,madelung,haug2008,vanderbilt} and reviews
\cite{sinitsyn2008}. Since the length-gauge time-dependent
Liouville equation has been a workhorse for  ultrafast dynamic
simulations from metals to insulators, it is imperative to provide a
resolution to this controversy.

In this paper, we show that the general Liouville equation in the
length gauge, even under a homogeneous electric field, has two partial
derivatives with respect to two different crystal momenta, in
agreement with Kohn and Luttinger's theory \cite{kohn1957}, but in
disagreement with Aversa and Sipe \cite{aversa1995}.  This
disagreement appears when the position operator is assumed to be
diagonal in $\bk$ \cite{aversa1995}, which leads to an extra Dirac
delta.  The fundamental reason is that the position operator does not
have a translational symmetry and has both $\bk$-diagonal and
$\bk$-off-diagonal elements. With the presence of an electric field,
the position operator couples the density matrix between two different
$\bk$.  As a result, the density matrix is a full matrix in $\bk$.
Our numerical test in a model semiconductor reveals that the
$\bk$-diagonal formalism produces a qualitatively different result and
underestimates both energy absorption and polarization by a factor of
7. This suggests that it become necessary to reexamine prior studies
based on Aversa and Sipe's equation \cite{aversa1995}.  Our study
develops the original idea of Kohn and Luttinger into a generalized
Liouville equation for the length gauge, where AS's $\bk$-diagonal
approach is a limiting with zero relative crystal momentum between two
states. This presents a general framework for dynamic simulations in
metals, semiconductors, insulators and other technologically advanced
materials.

The rest of the paper is arranged as follows. In Sec. II, we present
the theoretical formalism for a solid, and a detailed derivation of
the length-gauge time-dependent Liouville equation in the crystal
momentum representation. We will show our formalism agrees with KL's
original one, but not with AS' one.  Section III is on results and
discussions.  We conclude this paper in Sec. IV.


  \section{Theoretical formalism}

Since different notations were used in the literature
\cite{kohn1957,aversa1995}, we first define them consistently.  We
denote our Hamiltonian as $\hh$, which consists of two parts, $\hh_0$,
the unperturbed Hamiltonian, and $\hh_I$, the interaction between the
system and the external electric field, as \be
\hh=\hh_0+\hh_I=-\frac{\hbar^2\nab^2}{2m_e} +V(\br) + e\br \cdot
\cE(t), \ee where the first term is the kinetic energy operator, the
second is the potential with the full periodicity of the lattice, and
the last term is the interaction in the length gauge and does not have
the lattice periodicity due to the presence of $\br$. $\cE(t)$ is a
homogeneous external electric field that can be time-dependent.  Other
variables are standard. Without the external field, the system is
described by the Bloch wavefunction,
$\psi_n(\bkr)=u_n(\bkr)e^{i\bk\cdot\br}$, which satisfies the
eigenequation $\hh_0\psi_n(\bkr)=E_n(\bk)\psi_n(\bkr)$. Here
$E_n(\bk)$ is the $n$th band energy at crystal momentum $\bk$.
$u_n(\bkr)$ is the periodic part of the Bloch wavefunction, $
u_n(\bk,\br+\bR)=u_n(\bk,\br)$, and $\bR$ is a lattice vector.

The matrix elements of the position operator under the Bloch wave
representation are defined as $ \la \psi_n(\bkr)|\br|\psi_m(\bqr)\ra,$
where $\psi_n(\bkr)$ and $\psi_m(\bqr)$ are two Bloch states.
It has
been shown before \cite{epl25} that $\la
\psi_n(\bkr)|\br|\psi_m(\bqr)\ra$ is a full matrix in CMR, and has
both diagonal and off-diagonal elements in $n$ and $\bk$.
These elements can be computed directly \cite{epl25}, but a common
practice is to use the orthonormalization condition between the Bloch
wavefunctions \cite{blount1962,lax,matsyshya2019}. We follow Callaway
\cite{callaway} to denote two Bloch wavevectors by $\bk$ and $\bq$,
instead of $\bk$ and $\bk'$ used in \cite{aversa1995,kohn1957}, since
$\bk$ and $\bq$ are not equivalent. We start with the
orthonormalization condition $\la
\psi_n(\bkr)|\psi_m(\bqr)\ra=\delta(\bq-\bk)\delta_{nm}$, 
\footnote{\clr Writing the orthogonormalization as
$\delta(\bq-\bk)\delta_{nm}$ has a serious consequence. When one takes
the derivative of $\delta(\bq-\bk)$ with respect to either $\bq$ or
$\bk$, there is no derivative with respect to $n$ and $m$, which
restricts one to intraband transitions only (i.e., $n=m$) and narrows
the domain of the matrix elements of the position operator. This
difference can be seen by comparing the matrix elements of position
operator. The position operator is a full matrix in CMR among ($n,m$)
and ($\bk,\bq$), but taking the derivative amended with Berry
connection, one has either diagonal in $n$ only or diagonal in $\bk$
only, but no elements off-diagonal in both band index and crystal
momentum. In reality, $n(m)$ and $\bk(\bq)$ are inseparable since they
are a pair of indices for the same band state.  }
\be
\Omega^{-1} \int u_n^*(\bkr)u_m(\bqr)
\exp[i(\bq-\bk)\cdot \br] d\br =\delta(\bq-\bk)\delta_{nm},
\label{eq11}
\ee and take a partial-derivative of both sides with respect to $\bq$
to find for the left side to be \be \Omega^{-1} \int u_n^*(\bkr)u_m(\bqr)
i\br \exp[i(\bq-\bk)\cdot \br] d\br 
+ \Omega^{-1} \int
u_n^*(\bkr)[\nabla_\bq u_m(\bqr)]
\exp[i(\bq-\bk)\cdot \br] d\br, \label{eq10} \ee
and for the right side of Eq. \ref{eq11} to be
$[\nabla_\bq\delta(\bq-\bk)]\delta_{nm}$.
Since $\nabla_\bq\delta (\bq-\bk)$ is never equal to $\delta
  (\bq-\bk)$, the matrix element of the position operator must have
  $\bk$-off-diagonal elements.  The first term in Eq. \ref{eq10}
differs from the matrix element of the position operator by $i$, so we
divide both sides by $i$.  The second term in Eq. \ref{eq10} is
nonzero only if $\bk=\bq$ due to the lattice periodicity in
$u_n(\bkr)$ and $u_m(\bqr)$, which produces $\delta(\bq-\bk)$.  Doing
so, we have \be \la \psi_n(\bk,\br) |\br |\psi_m(\bq,\br)\ra =-i
[\nabla_\bq\delta (\bq-\bk)] \delta_{n,m}
+\delta(\bq-\bk)\frac{i}{\Omega} \int_\Omega u^*_n(\bk,\br) \nabla_\bq
u_m(\bq,\br) d\br, \label{u5} \ee which exactly matches Blount's
equation 2.3 \cite{blount1962}.  Equivalently, one may take the
derivative of Eq. \ref{eq11} with respect to $\bk$, \be \la
\psi_n(\bk,\br) |\br |\psi_m(\bq,\br)\ra =+i [\nabla_\bk \delta
  (\bq-\bk)] \delta_{n,m}         -\delta(\bq-\bk)\frac{i}{\Omega}
\int_\Omega [\nabla_\bk u^*_n(\bk,\br)] u_m(\bq,\br) d\br. \label{u6}
\ee One sees clearly that although the final result must be the same,
two terms separately in Eq. \ref{u6} are different from those in
Eq. \ref{u5}, so they must appear in pair \cite{aversa1995}.  We shall
work with Eq. \ref{u5}, that is, the partial derivative always applies
to $\psi_m(\bqr)$ on the right side of the position operator $\br$ in
Eq. \ref{u5}.  We denote the last term behind $\delta(\bq-\bk)$ in
Eq. \ref{u5}, i.e., the Berry connection, as $
\bj(n\bk;m\bq)=\frac{i}{\Omega} \int_\Omega u^*_n(\bk,\br) \nabla_\bq
u_m(\bq,\br) d\br, $  so Eq. \ref{u5} becomes \be \la
\psi_n(\bk,\br) |\br |\psi_m(\bq,\br)\ra =-i [\nabla_\bq \delta
  (\bq-\bk)] \delta_{n,m} +\delta(\bq-\bk) \bj(n\bk;m\bq),
       \label{u7} \ee
which will be used repeatedly below.  KL's $J$ in Eq. \ref{kohn1}
\cite{kohn1957} is restricted to  $\bk$, not generic enough
for our purpose.



\newcommand{\hpsi}{\hat{\psi}}
\newcommand{\has}{\hat{a}}
\newcommand{\bbx}{{\bf r}}

The time-dependent Liouville equation is deeply rooted in the
Heisenberg equation of motion for an operator. It is helpful to review
some key steps that lead to the Liouville equation within the
many-electron Hamiltonian. We
follow Ehrenreich and Cohen (EC) \cite{ehrenreich1959} and pages
101-103 \cite{kittel3}  and
introduce a field operator $\hpsi$ in the coordinate representation
 defined as $\hpsi(\bbx)=\sum_\bk\has_\bk e^{i\bk\cdot \bbx}$,
where $\has_\bk$ is the annihilation operator, $\bbx$ is the coordinate
and $\bk$ is the crystal momentum. The time evolution of $\hpsi(\bbx)$
is $i\hbar\frac{\partial \hpsi}{\partial t}=[\hpsi,\hh(\br)]$. The density
matrix operator is  $\hr(\bbx,\bbx')=\hpsi^\dag(\bbx') \hpsi(\bbx)$,
then we have the Liouville equation in the $\bbx$ representation,
\be
i\hbar \dot{\hr}(\bbx,\bbx')=\hh(\bbx)\hr(\bbx,\bbx')-
\hr(\bbx,\bbx')\hh(\bbx'),
\label{eq601}
\ee
where $\hr$ depends on two spatial points, $\bbx$ and $\bbx'$. In LG,
the Hamiltonian is 
$\hh(\br)=\hT(\br)+V(\br)+e\br\cdot \cE(t)$, where $\hT$ is the
kinetic energy operator, $V(\br)$ is the lattice potential and
$e\br\cdot\cE(t)$ is the interaction between the light field and
system.
We can also transform $\hr(\bbx,\bbx')$ to 
 the $\bk$ representation, \be \hr(\bbx,\bbx')
 =\frac{1}{\Omega}\sum_{\bk,\bk'}\hr(\bk,\bk')e^{i(\bk\cdot
   \bbx-\bk'\cdot \bbx')},\ee whose inverse Fourier transform is 
\be
\hr(\bk,\bk')=\int\int d\br d\br'\hr(\bbx,\bbx')e^{-i\bk \cdot
  \br+i\bk'\cdot \br'}.
\ee
There is no symmetry between $\bk$ and $\bk'$ since they are
independent of each other. One can think of $\bk$ and $\bk'$
representing the crystal momentum of a hole and that of an electron,
respectively \cite{mahan2000}. This asymmetry can be found in the
literature as well. 
For the Coulomb potential, EC obtained
(their equation 17), \ba i\hbar (\partial/\partial
t)\hr(\bk,\bk+\bq)&=&(E_\bk-E_{\bk+\bq})\hr(\bk,\bk+\bq)\nonumber\\ &+&\sum_{\bk',\bq'}v_{\bq'}[
  \hr(\bk',\bk'+\bq')\hr(\bk+\bq',\bk+\bq)-
  \hr(\bk,\bk+\bq-\bq')\hr(\bk',\bk'+\bq'), \nonumber
    \ea
where the density matrix operator depends on  two crystal momenta in
the $\bk$ representation.

Going from the $\br$ representation in Eq. \ref{eq601} to the Bloch
state representation with $|n\bk_1\ra$ and $|m\bk_2\ra$ as the basis functions is
straightforward,  \ba i\hbar \dot{\rho}(n\bk_1;m\bk_2) &=& \la
n\bk_1|[\hh_0,\hr]|m\bk_2\ra+\la n\bk_1|[\hh_I,\hr]|m\bk_2\ra\breaker
&=&(E_n(\bk_1)-E_m(\bk_2)) \rho(n\bk_1;m\bk_2) +\la
n\bk_1|(\hh_I\hr- \hr \hh_I)|m\bk_2\ra.
\label{eq20}
\ea Our goal
is to investigate whether the position
operator could lead to a diagonal term in the second line in
Eq. \ref{eq20}, i.e. $\bk_1=\bk_2$.  We adopt a continuous $\bk$, and
insert a completeness $\int d\bq \sum_l |l\bq \ra\la l\bq|=1$ between
$\hh_I$ and $\hr$ in Eq. \ref{eq20}, \ba i\hbar
\dot{\rho}(n\bk_1;m\bk_2)&=&
(E_n(\bk_1)-E_m(\bk_2))\rho(n\bk_1;m\bk_2)\nonumber \\& +&\sum_l\int
d\bq \left ( \la n\bk_1|\hh_I |l\bq\ra\la l\bq|\hat{\rho}|m\bk_2\ra
- \la n\bk_1|\hat{\rho} |l\bq\ra\la
l\bq|\hh_I|m\bk_2\ra\right ) \breaker
&=&(E_n(\bk_1)-E_m(\bk_2))\rho(n\bk_1;m\bk_2)
+e\cE(t)\cdot
\sum_l\int d\bq \la n\bk_1|\br|l\bq\ra \rho(l\bq;m\bk_2) \breaker
&-&e\cE(t)\cdot \sum_l \int d\bq \rho(n\bk_1;l\bq) \la l\bq
|\br|m\bk_2\ra
\label{eq12}, \ea
where we have used $\hh_I=e\br \cdot \cE(t)$.  This equation is
generic, where two terms, lines 2 and 3 in Eq. \ref{eq12}, have the
position operator, but Kohn and Luttinger \cite{kohn1957} did not work
with those singular derivatives associated with the position operator;
instead they introduced a specialized Hamiltonian and then reduced the
commutator ${\cal C}$ to a limit.  Here, we introduce a
straightforward method.

We notice  that $\la n\bk_1|\br|l\bq\ra$ and $ \la l\bq
|\br|m\bk_2\ra$ are structurally different, which is often ignored in
standard textbooks \cite{kittel3}.
We first treat $\la n\bk_1|\br|l\bq\ra$ in Eq. \ref{eq12}.  Using
Eq. \ref{u7}, we can rewrite the first line in Eq. \ref{eq12} to \be
e\cE(t)\cdot \sum_l \int d\bq [ -i [\nabla_\bq\delta (\bq-\bk_1)]
  \delta_{n,l}+\delta(\bq-\bk_1) \bj(n\bk_1;l\bq)
]\rho(l\bq;m\bk_2).
\label{eq14}
\ee A key feature of Eq. \ref{eq14} is that the nabla operator
$\nabla$ and the Dirac delta $\delta$ are intertwined.  We realize
that the first term in Eq. \ref{eq14} needs the help of the identity,
$\int \frac{\partial \delta(y-x)}{\partial y} f(y)dy
=-\frac{\partial}{\partial x}f(x)$.  This identity is conditional:
$f(y)$ must have the same value in the lower and upper limits, or is
periodic.  $\rho(l\bq;m\bk_2)$ is indeed periodic in $\bq$ from the
Brillouin boundary to another boundary, $\rho(l\bq+\bG;m\bk_2)=
\rho(l\bq;m\bk_2)$. Note that this periodicity refers to the
reciprocal lattice periodicity, not the real space translational
invariance.
So we can simplify Eq. \ref{eq14} 
to
\be
e\cE(t)\cdot 
 [i 
\nabla_{\bk_1}\rho(n\bk_1;m\bk_2)
+\sum_l \bj(n\bk_1;l\bk_1)
\rho(l\bk_1;m\bk_2)     ].
\label{eq15}
\ee

Next, we treat the second line in Eq. \ref{eq12} similarly as the
first line, but there is a difference.  The partial derivative is
taken with respect to $\bk_2$, not to $\bq$.  \Ba &-\sum_l \int d\bq
\la n\bk_1|\hat{\rho} |l\bq\ra\la l\bq|\hh_I|m\bk_2\ra =-e\cE(t)\cdot
\sum_l \int d\bq \rho(n\bk_1;l\bq) \la l\bq |\br|m\bk_2\ra \breaker
&=-e\cE(t)\cdot \sum_l \int d\bq \rho(n\bk_1;l\bq) \left ( -i
    [\nabla_{\bk_2}\delta (\bk_2-\bq)] \delta_{l,m} +\delta(\bk_2-\bq)
    \bj(l\bq;m\bk_2) \right ) \breaker &=e\cE(t)\cdot [
      i\nabla_{\bk_2}\rho(n\bk_1;m\bk_2) -\sum_l
      \rho(n\bk_1;l\bk_2)\bj(l\bk_2;m\bk_2) ].
\label{eq18}
\end{align}
  In the last step since 
 the derivative is taken
 over $\bk_2$, not $\bq$ as Eq. \ref{eq14}, we have
  rewritten 
$\int \frac{\partial }{\partial x}\delta(x-y)f(y)dy$ 
 as  $ \frac{\partial }{\partial x} \int 
\delta(x-y)f(y)dy =\frac{\partial }{\partial x} f(x)$ before we 
carry out the integration.

Finally,  we collect Eqs. \ref{eq15} and   \ref{eq18}, together 
with  the first term in
Eq. \ref{eq12} to obtain
\ba
i\hbar
\dot{\rho}(n\bk_1;m\bk_2) &=&
(E_n(\bk_1)-E_m(\bk_2))\rho(n\bk_1;m\bk_2)
+ie\cE(t)\cdot 
\left 
  (
  \nabla_{\bk_1}
  +\nabla_{\bk_2}
\right ) 
 \rho(n\bk_1;m\bk_2)
\breaker
&+&e\cE(t)\cdot  [
\sum_l \bj(n\bk_1;l\bk_1)
\rho(l\bk_1;m\bk_2)
-\rho(n\bk_1;l\bk_2)\bj(l\bk_2;m\bk_2)  ],
\label{eq30}
\ea 
 where the last two terms exactly match Eq. 113 of Kohn and Luttinger
 \cite{kohn1957}, without their complicated procedure.  Equation
 \ref{eq30} shows that there are two partial derivatives with respect
 to two crystal momenta, $\bk_1$ and $\k_2$. We can now explain why
 Eq. \ref{eq30} is reasonable physically.  Because the commutator,
 $[\br,\hr]=\br \hr- \hr\br$, has a symmetry between $\br$ and $\hr$,
 when we compute its matrix elements between $\la n\bk_1|$ and
 $|m\bk_2\ra$, one position operator $\br$ is close to $\bk_1$ of
 $|n\bk_1\ra$ and the other $\br$ is close to $\bk_2$ of
 $|m\bk_2\ra$. Since each position operator is associated with the
 $\nabla \delta(\bk_1-\bk_2)$, if there is a partial derivative with
 respect to $\bk_1$, there must be also another one with respect to
 $\bk_2$, yielding an intrinsic symmetry between $\bk_1$ and
 $\bk_2$. This explains why we have two partial derivatives, not one.
 Since $\hr$ always goes together with $\br$, band states $|n\bk_1\ra$
 and $|m\bk_2\ra$ are always separated from each other by either
 $\hr\br$ or $\br\hr$ and cannot be contracted to a single Dirac
 $\delta(\bk_1-\bk_2).$ This unambiguously proves that KL's equation
 is correct.

\section{Results and Discussions}

We now examine the work by Aversa and Sipe
  \cite{aversa1995}. They did not provide the details of their
  derivation, but it is not too difficult to pin down the root cause
  of the discrepancy.

\subsection{$\bQ=0$ limit}

  It is possible to further simplify Eq. \ref{eq30} by converting two
 partial derivatives to a single derivative by properly changing
 arguments \cite{bufler1994,hess1996}. We set $\bk_1=\bq+\bQ$ and
 $\bk_2=\bq-\bQ$, so $\bq=(\bk_1+\bk_2)/2$,
 $\bQ=(\bk_1-\bk_2)/2$. Then two partial derivatives become
 $\frac{\partial }{\partial \bk_1}= \frac{\partial \bq}{\partial
   \bk_1} \frac{\partial }{\partial \bq} + \frac{\partial
   \bQ}{\partial \bk_1} \frac{\partial }{\partial \bQ}=
 \frac{1}{2}\frac{\partial }{\partial \bq}+ \frac{1}{2}\frac{\partial
 }{\partial \bQ} $ and $\frac{\partial }{\partial \bk_2}=
 \frac{\partial \bq}{\partial \bk_2} \frac{\partial }{\partial \bq} +
 \frac{\partial \bQ}{\partial \bk_2} \frac{\partial }{\partial \bQ}=
 \frac{1}{2}\frac{\partial }{\partial \bq}- \frac{1}{2}\frac{\partial
 }{\partial \bQ} $, so $\frac{\partial }{\partial
   \bk_1}+\frac{\partial }{\partial \bk_2}=\frac{\partial }{\partial
   \bq}$. Then Eq. \ref{eq30} becomes
 \begin{flalign}
 &i\hbar
    \dot{\rho}(n,\bq+\bQ;m,\bq-\bQ) \breaker&=
    (E_n(\bq+\bQ)-E_m(\bq-\bQ))\rho(n,\bq+\bQ;m,\bq-\bQ) 
    +ie\cE(t)\cdot \nabla_\bq \rho(n,\bq+\bQ;m,\bq-\bQ) \breaker
    &+e\cE(t)\cdot \sum_l \bj(n,\bq+\bQ;l,\bq+\bQ)
    \rho(l,\bq+\bQ;m,\bq-\bQ) \breaker
    &-\rho(n,\bq+\bQ;l,\bq-\bQ)\bj(l,\bq-\bQ;m,\bq-\bQ),
\label{eq31}
\end{flalign}
 where $\bq$ and $\bQ$ are two independent crystal momenta. $\bq$
represents the center momentum while $\bQ$ is their relative
one. Equation \ref{eq31} does not change the nature of our conclusion,
but it does allow us to find some limiting cases easily. 
Setting $\bQ=0$ yields
\ba
i\hbar
\dot{\rho}(n,\bq;m,\bq) =
(E_n(\bq)-E_m(\bq))\rho(n,\bq;m,\bq)
+ie\cE(t)\cdot 
\nabla_\bq
 \rho(n,\bq;m,\bq)\breaker
+e\cE(t)\cdot [ 
\sum_l \bj(n,\bq;l,\bq)
\rho(l,\bq;m,\bq)
-\rho(n,\bq;l,\bq)\bj(l,\bq;m,\bq) ],
\label{eq32}
\ea
which reproduces AS' 
Eq. 9 \cite{aversa1995}. This shows that their equation is really a
limiting case with $\bQ=0$.




\subsection{Will the $\bQ=0$'s density matrix time-evolve to $\bQ\ne 0$ one?}

Since AS' equation has a partial derivative of $\rho$ with respect to
$\bk$, i.e., $i\frac{\partial \rho}{\partial \bk}$, will this be
enough to generate the missing coupling of $\rho$ between different
$\bk$ points?  Numerically, the partial derivative is replaced by a
finite difference \cite{virk2007}. We choose three $k$ points along a
single straight line in the $\bk$ space, $k_1,k_2,k_3$, with the
minimum difference denoted as $\delta k$, $i\frac{\partial
  \rho}{\partial \bk} \approx i \frac{\rho(k+\delta k)-\rho(k-\delta
  k)}{2\delta k}.  $ We limit ourselves to a single band.
$i\frac{\partial \rho}{\partial \bk}$ is represented by a $3\times 3$
matrix as \ba i\frac{\partial \rho(k,k)}{\partial \bk} \approx i\gamma
\begin{pmatrix}
  0 & \rho(k_2,k_2) & 0 &\\
  -\rho(k_1,k_1) &0 &\rho(k_3,k_3) \\
  0 &-\rho(k_2,k_2)& 0 \\
\end{pmatrix}, \nonumber
\ea where $\gamma=\frac{1}{2\delta k}$.  One sees that all the density
matrices are $\bk$-diagonal. Only the density matrices at $k_i$ and
$k_{i\pm 1}$ are connected to each other, like between $k_1$ and
$k_2$, but not between $k_1$ and $k_3$. This is very different from
the viewpoint of Eq. \ref{eq12}.

In Eq. \ref{eq12}, the position operator has both $\bk$-diagonal and
$\bk$-off-diagonal, so the density matrix is a full matrix \be
\begin{pmatrix}
  \rho(k_1,k_1) & \rho(k_1,k_2)      & \rho(k_1,k_3) \\
  \rho(k_2,k_1)       &\rho(k_2,k_2) &\rho(k_2,k_3)      \\
  \rho(k_3,k_1) &\rho(k_3,k_2)& \rho(k_3,k_3)\\
\end{pmatrix}. \nonumber
\ee Because $\br$ couples different $\bk$, even if initially the
off-diagonal matrix elements of the density matrix are zero, they will
become nonzero after we turn on the electric field. Both the uniform
and nonuniform electric fields couple different $\bk$, because they
both have $\br$.  For these reasons, AS' scheme is incomplete since it
is restricted to $\bQ=0$, i.e. $\bk$-diagonal one only.  Our finding
will affect all the prior results such as
\cite{golde2008,pedersen2015} that used their equation (Eq. 20).
{\clr Below, we will provide several numerical examples for this
and also in the Supplementary Material.}

\subsection{A wavefunction perspective}

The density matrix is a product of two wavefunctions.  In the
unperturbed system, it is $\psi_n^*(\bk_1,\br)\psi_m(\bk_2,\br)
=e^{i(\bk_2-\bk_1)\cdot \br} u_n^*(\bk_1,\br) u_m(\bk_2,\br) $. If we
change our notations by setting $\bk_2-\bk_1=\bq$ and
$u_n^*(\bk_1,\br) u_m(\bk_2,\br)=u_{\nu}(\bq,\br)$, we see the density
matrix is like another Bloch wavefunction which is diagonal in
$\bq$. But as soon as we turn on $ \hH_I(\br)$, it breaks the
translational symmetry $\hH_I(\br+\bR_l) \ne \hH_I(\br)$, then
$\psi_n(\bkr)$ is not an eigenfunction anymore. The actual
eigenfunction is a linear combination of those Bloch wavefunctions,
\be \phi_1(\br)=\sum_{n\bk_1}C_{n\bk_1}\psi_n(\bk_1,\br),
\hspace{0.2cm} \phi_2(\br')=\sum_{m\bk_2}C_{m\bk_2}\psi_m(\bk_2,\br'),
\ee and the density matrix is now \be
\phi_1^*(\br)\phi_2(\br')=\sum_{n\bk_1;m\bk_2}C_{n\bk_1}^*
C_{m\bk_2}\psi_n^*(\bk_1,\br)\psi_m(\bk_2,\br'), \ee which makes it
clear that the density matrix in general is a full matrix in $\bk$.

\subsection{An analytical example}

Since our paper is mainly analytical, an extensive numerical test in
different real systems is beyond the scope of this paper, and is left
for future investigations. Here, we limit ourselves to some simple
models that the reader can easily verify.

To convince ourselves that an electric field in the length gauge
  destroys the periodicity of the charge density, we employ a two-site
  model, with the Hamiltonian \be \hh=\begin{pmatrix} -e\cE a/2 & -t
  \\ -t & e\cE a/2 \\
\end{pmatrix},
\ee where $\cE$ is the electric field, and $a$ is the lattice
constant. We diagonalize it to obtain the eigenenergies
$\epsilon_{1,2}=\pm \sqrt{t^2+(e\cE a/2)^2}$. The charge densities at
sites 1 and 2 for the state with eigenenergy of $\epsilon_1$ are $
\rho_1=\frac{t^2}{t^2+(\epsilon_1+e\cE a/2)^2};
\rho_2=\frac{(\epsilon_1+e\cE a/2)^2}{t^2+(\epsilon_1+e\cE a/2)^2}, $
which proves clearly that the charge density with the presence of the
electric field is not periodic.

\subsection{A  metal}

We consider a model with three bands \cite{souza2002,souza2004}, \be
\hh=-t_0\sum_{l} (\hcp_{l}\hc_{l+1}+ {\rm h.c.})-V_0\sum_l \cos[2\pi
  la/(3a)+2\pi/3] \hcp_{l}\hc_{l} , \ee where $l$ runs over all the
lattice sites, $V_0$ is the site energy and controls the energy gap,
$a_0=3a$ is the lattice constant and other parameters have the same
meaning as above.   {\clr The hopping integral $t_0$
  determines the band width and transition.  Figure \ref{fig1}(a)
  illustrates our model to simulate either a metal or a semiconductor
  by using different $t_0$ and fillings, where each unit cell contains
  three atoms. The laser impinges down with the electric field along
  the chain direction. On the left of Fig. \ref{fig1}(b) is a typical
  metal band structure, where the lowest band is concave up and
  disperses like a nearly free electron model. For this reason, we
  select $a=1\rm\AA$, $t_0=1$ eV and $V_0=1$ eV.  We choose 81 sites
  with the periodic boundary condition.  Diagonalizing the Hamiltonian
  matrix gives us the band structure as shown in Fig. \ref{fig2}(a),
  which matches our expectation in Fig. \ref{fig1}(b) well.}  The
total number of lattice sites is 81, so in the reciprocal space, the
Brillouin zone has 27 points, $k_i(a_0)=\frac{\pi i}{13}$, where $i$
runs from $-13$ to 13.  We fill one-third of the lowest band states,
so we have a metal. These eigenstates allow us to compute the
transition matrix elements for the position operator (see the details
in \cite{epl25}), without converting the position operator to the
partial-derivative with respect to $\bk$ as done above, since the
latter is prone to numerical errors and becomes even harder when bands
are degenerate.  Then we can solve the time-dependent Liouville
equation numerically, $i\hbar \dot{\rho}(n\bk_1;m\bk_2) = \la
n\bk_1|[\hh_0,\hr]|m\bk_2\ra+\la n\bk_1|[\hh_I,\hr]|m\bk_2\ra
=(E_n(\bk_1)-E_m(\bk_2)) \rho(n\bk_1;m\bk_2)+ \sum_{l\bq} \la
n\bk_1|\hh_I|l\bq\ra\la l\bq|\hr|m\bk_2\ra- \la n\bk_1|\hr|l\bq\ra\la
l\bq| \hh_I|m\bk_2\ra.$ We choose a 20-THz-laser pulse of 100 fs and
electric field of 0.01 $\rm V/\AA$.  Figure \ref{fig2}(b) shows the
time-dependent real (solid line) and imaginary parts (dashed line) of
the density matrix $\rho(k_0,k_4)$ between two band states at two
different crystal momenta, $k_0$ and $k_4$, where
$k_ia_0=\pi\frac{i}{13}$.  It is clear that off-diagonal matrix
elements are nonzero, though initially they are zero.

\begin{figure}
  \includegraphics[angle=0,width=1\columnwidth]{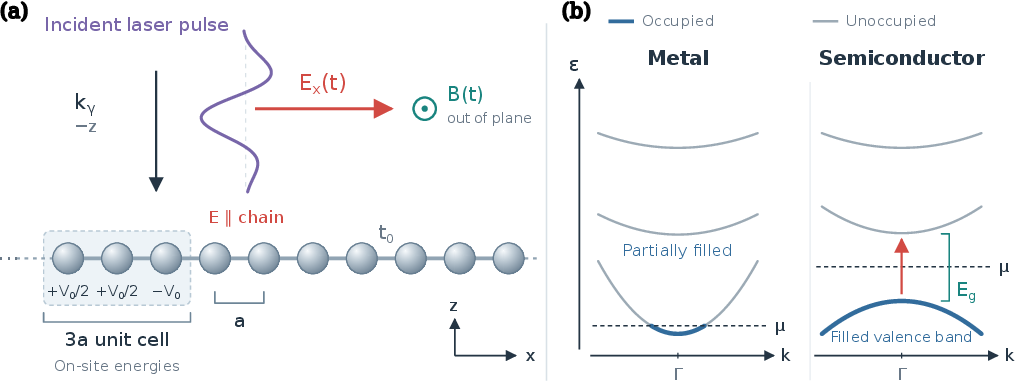}
  \caption{(a) Laser-induced optical excitation in our model system
    which is used to simulate both a metal and semiconductor, but with
    different hoppings and fillings.  (b) (Left) A typical band
    structure of a metal, where the lowest band is concave up. (Right)
    A semiconductor band structure. The lowest band is concave
    down. The arrow represents the excitation.  }
\label{fig1}
\end{figure}

\begin{figure}
  \includegraphics[angle=0,width=0.7\columnwidth]{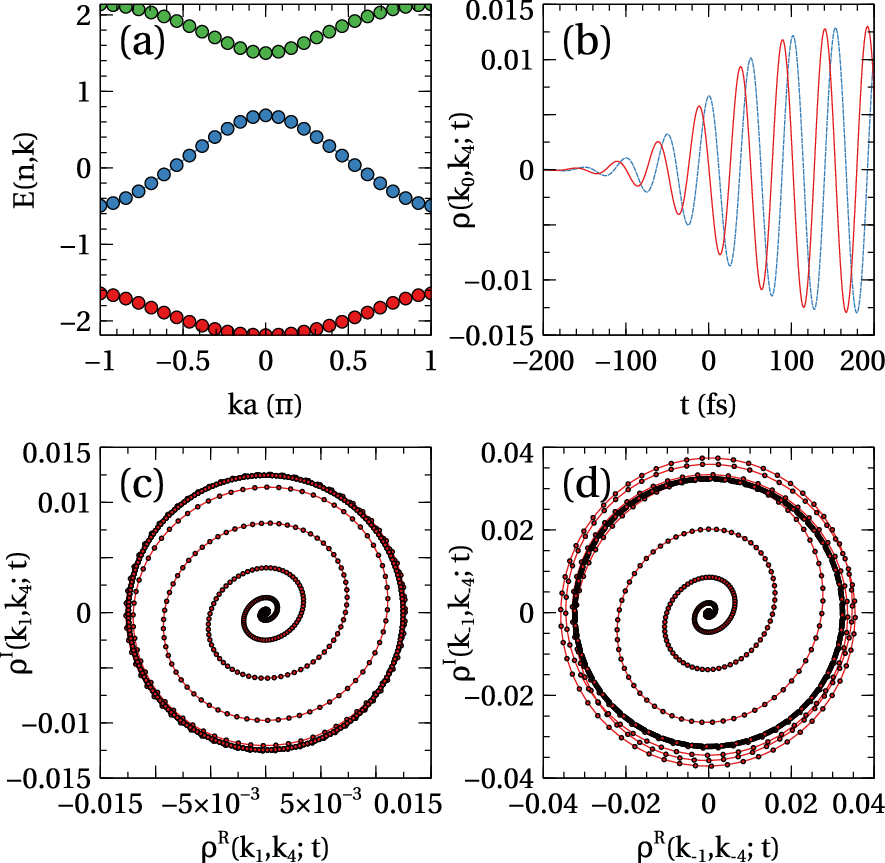}
  \caption{ (a) Band structure. (b) Density matrix $\rho(k_0,k_{4})$ as
    a function of time. The solid line denotes the real part and the
    dashed line denotes the imaginary part. (c) Imaginary part versus
    the real part of $\rho(k_1,k_4)$. (d) Imaginary part versus the
    real part of $\rho(k_{-1},k_{-4})$.
  }
\label{fig2}
\end{figure}

Other pairs of density matrices show a similar feature.  Figure
\ref{fig2}(c) plots the imaginary part of the density matrix
$\rho(k_1,k_4)$ versus the real part. This allows us to see their
phase relation. They start from the center and form a spiral
pattern. A similar pattern is also found for $\rho(k_{-1},k_{-4})$,
see Fig. \ref{fig2}(d). Using a larger system or changing the band
filling lead to the same conclusion.  This proves that within the
length gauge, density matrices are coupled between two different $\bk$
points. This is consistent with our analytic expression derived above.

\subsection{A  semiconductor}

\begin{figure}
  \includegraphics[angle=0,width=0.7\columnwidth]{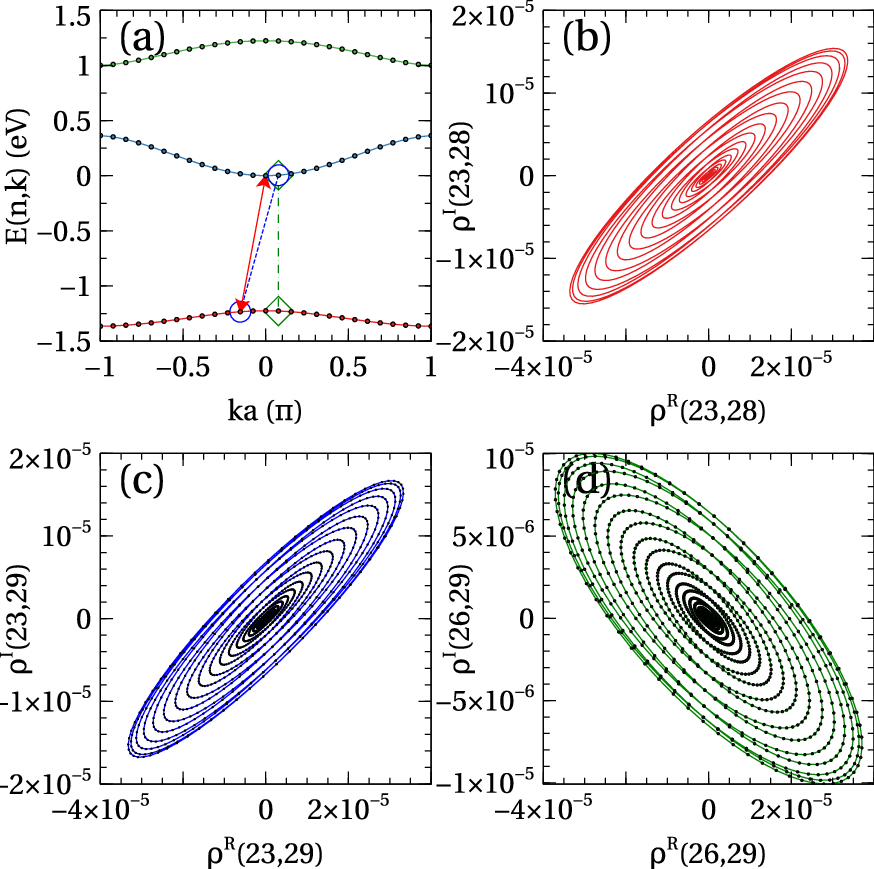}
  \caption{(a) Band structure of a semiconductor.  (b) Imaginary part
    versus the real part of $\rho(23,28)$.  (c) Imaginary part versus
    the real part of $\rho(23,29)$.  (d) Imaginary part versus the
    real part of $\rho(26,29)$.  Here states 23, 26, 28 and 29 are
band 1 at $ka_0=-\frac{2}{13}\pi$, band 1 at $ka_0=\frac{1}{13}\pi$, 
band 2 at $ka_0=\frac{0}{13}\pi$,  and  band 2 at $ka_0=\frac{1}{13}\pi$,
respectively. 
}
\label{fig3}
\end{figure}

Next, we consider a semiconductor. It is necessary to make several
changes. {\clr Figure \ref{fig1}(b) (the right figure) shows
  that semiconductors have a lowest band concave down. Therefore, we
  choose $t_0=-0.5$ eV, so the lowest band in the band structure is
  concave down, similar to a typical semiconductor.  We still keep
$V_0=1$ eV, but we fill the lowest band, so the band gap appears between
the valence band and conduction band at the $\Gamma$ point
$ka_0=0$. Figure \ref{fig3}(a) is our band structure.  With our present
parameters, the band gap is $E_g= 1.224$ eV, which is reasonable for a
model semiconductor. }

Another thing that is different from a metal is that semiconductors
mainly rely on the interband transition since the lower bands are
completely filled. {\clr We use a 50-THz pulse with the field
  amplitude of 0.01 $\rm V/\AA$ and duration of 100 fs.  Figure
  \ref{fig3}(b) shows the imaginary part versus the real part of the
  density matrix between two bands at two different crystal momenta,
  forms an ellipse, different from our model metal above. The real
  part of the density matrix is larger than the imaginary part.  The
  trajectory of the density matrix element is in the first and third
  quadrants.  Number 23 in the figure refers to the state in the
  lowest band at $ka_0=-\frac{2}{13}\pi$ and 28 is in the second band at
  $ka_0=0.0000$.  This transition is highlighted by the red
  double-arrowed line in Fig. \ref{fig3}(a). Figure \ref{fig3}(c)
  shows the density matrix element $\rho(23,29)$ between states 23 and
  29, with their transition highlighted by the blue double-circled
  line in Fig. \ref{fig3}(a).  Figure \ref{fig3}(d) shows the results
  for the vertical transition between states 26 and 29.  This
  transition is denoted by the dark green double-boxed line in
  Fig. \ref{fig3}(a).  The trajectory of the density matrix element is
  in the second and fourth quadrants.} We find the results are similar
regardless of the system size and filling, {\clr see our
  supplementary materials}. Within the limitation of our model, we can
safely conclude that within the length gauge, the density matrices at
different $\bk$ points are coupled.

\begin{figure}
  \includegraphics[angle=0,width=0.7\columnwidth]{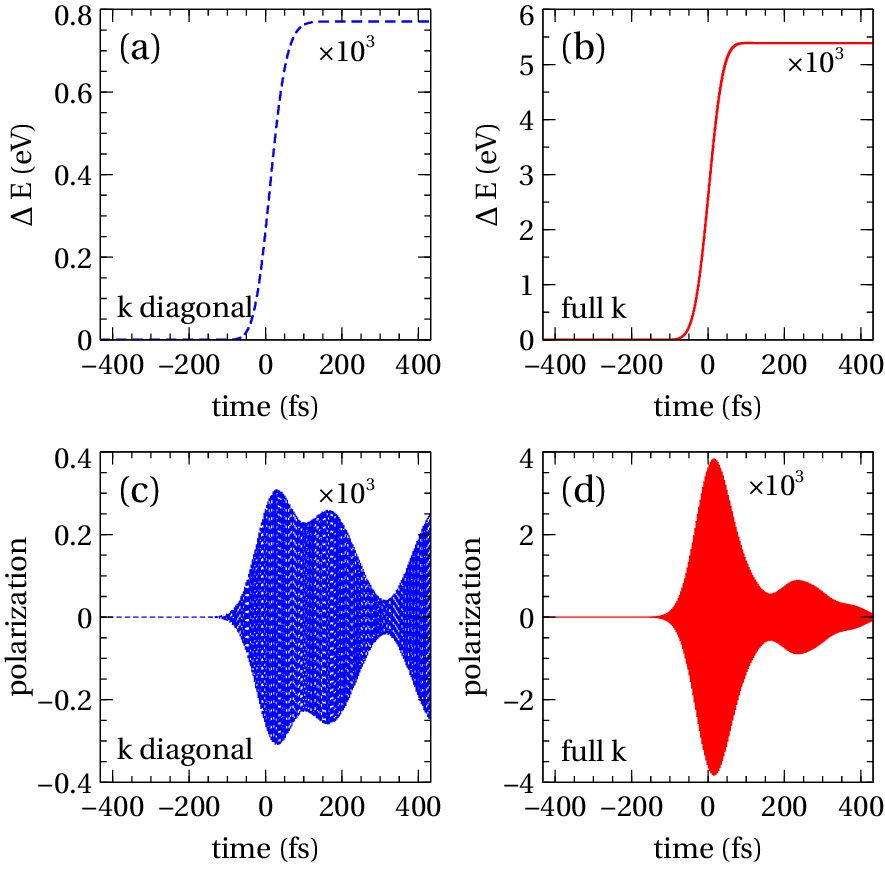}
  \caption{ (a) Energy absorbed into the system as a function of time,
    where the  $\bk$-diagonal formalism is used. Note the data is
    multiplied by $10^{3}$ for an easy view.  (b) Same as (a) but
    using our full $\bk$ formalism, where both $\bk$-diagonal and
    $\bk$-off-diagonal elements are included in the transition matrix
    elements. (c) Polarization as a function of time computed with
     $\bk$-diagonal formalism. The data is also multiplied by
    $10^{3}$. (d) Same as (c) but using our full $\bk$ formalism.  }
\label{fig4}
\end{figure}

Finally, we wish to quantify the difference between the AS'
$\bk$-diagonal formalism and our full $\bk$ formalism. Two quantities
are computed. One is the total energy absorbed into the system
($\Delta E(t)=E(t)-E(0)$) and the other is the induced polarization
($P(t)=-e{\rm Tr}(\rho\br)$), where Tr denotes the trace. We use the
same initial condition of the system (i.e., the same number of $\bk$
points and the same band filling), and keep the rest of the laser
parameters as above, except the laser photon energy changed to 1.22 eV
to have a real excitation.  The results using AS formalism are given
in Figs. \ref{fig4}(a) and \ref{fig4}(c), while those using our
formalism in Figs. \ref{fig4}(b) and (d).  Figure \ref{fig4}(a) shows
that the total energy absorbed into the system increases quickly upon
laser excitation and reaches 0.8 meV. Note that all the data in
Fig. \ref{fig4} is multiplied by 1000 for an easy view. Figure
\ref{fig4}(b) shows that the absorbed energy $\Delta E$ reaches 5.5
meV, about 7 times larger than the AS formalism.

A much larger difference is found for polarization.  The polarization
in Fig.  \ref{fig4}(c) oscillates rapidly due to the higher photon
energy, and has a double hump within 200 fs. The polarization computed
within our formalism is both qualitatively and quantitatively
different.  Figure \ref{fig4}(d) has a single big peak, slightly after
0 fs. The amplitude is also about 7 times larger.  There is no
similarity between AS and our results. This shows that
$\bk$-off-diagonal elements of $\rho$ have a significant effect on the
final results.  {\clr In our Supplementary Material,} we provide
extensive tests, including the convergence tests on the system size,
laser pulse duration and photon energies. All of them reach the same
conclusion that the $\bk$-diagonal Liouville equation systematically
underestimates the polarization.  So far, our test is limited to two
model systems. Further research is necessary to see whether our
conclusion remains valid in real materials.  

\subsection{\clr A system with open boundary condition}

  \begin{figure}
  \includegraphics[angle=0,width=0.7\columnwidth]{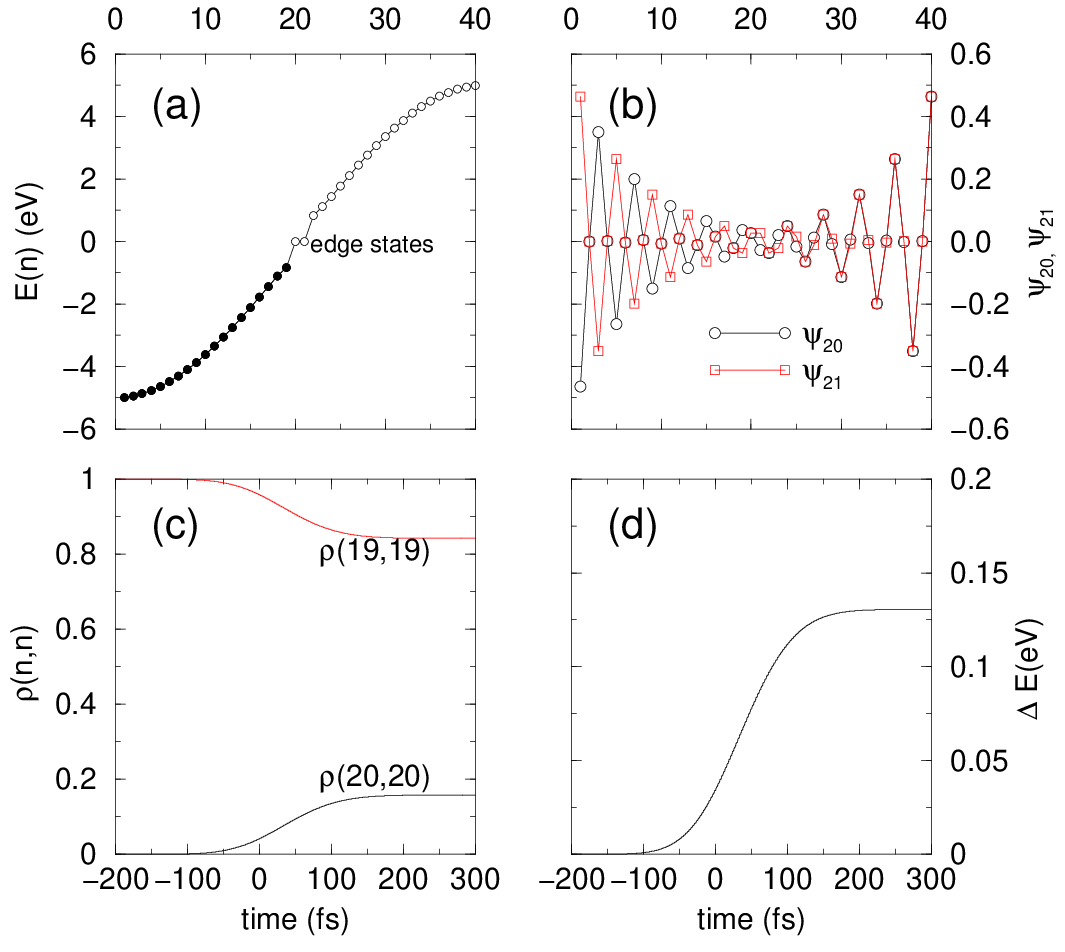}
  \caption{\clr (a) Eigenenergies as a function of energy index for a
    chain of 40 sites with open boundary condition, where two edge
    states are highlighted.  (b) The wavefunctions of two edge states,
    states 20 and 21.  (c) Time evolution of the populations in states
    19 and 20. (d) Energy absorbed ($\Delta E$) as a function of time.}
\label{fig8}
\end{figure}

{\clr Since most of technological advanced materials are quantum
  confined, to show that our formalism is also versatile for such
  systems, we adopt Su-Schrieffer-Heeger (SSH) model
  \cite{su1979}. The model describes a quasi-one-dimensional
  polyacetylene. Due to the Peierls distortion, the carbon chain is
  dimerized, with short and long bonds alternating along the chain, so
  there are two atoms per unit cell. The Hamiltonian is, \be
  \hh=-\sum_l [t_0+(-1)^lt_1]\hcp_{l}\hc_{l+1}+h.c., \ee where $t_0=2.5$
  eV and $t_1=0.35$ eV. As an example, we choose $N=40$ sites with an
  open boundary condition, so we cannot use $k$ to label states
  anymore.  Once we diagonalize the Hamiltonian matrix, we obtain the
  eigenvalues which are shown in Fig. \ref{fig8}(a). We see that
  besides a quasi-dispersion similar to the periodic boundary
  condition case \cite{su1979}, two edge states appear.  Their
  respective wavefunctions $\psi_{20}$ and $\psi_{21}$
  in real space are illustrated in
  Fig. \ref{fig8}(b). It is clear that these edge states are mainly
  confined to the ends of the chain.

 With these wavefunctions in hands, we can compute the matrix elements
 of the position operator between eigenstates. We assume that all 19
 energy levels are occupied (see the filled circles in
 Fig. \ref{fig8}(a)), with the rest unoccupied. We tune the laser
 photon energy to match the energy difference between the top of the
 valence band, state 19, and the first edge state, state 20, i.e.,
 $h\nu=0.826$ eV.  The electric field amplitude is $\cE_0=0.05 \rm
 V/\AA$, and the duration is $\tau=100$ fs. We solve the
 time-dependent Liouville equation similarly as above.  Figure
 \ref{fig8}(c) shows how the populations in states 19 and 20 change
 with time. In the beginning, state 19 is completely occupied with the
 occupation being 1, while that in state 20 is 0. As time progresses,
 the laser pulse excites the electron from state 19 to 20 by 0.16
 electron. This allows the system to absorb the energy from the laser
 field. The energy change with time is shown in
 Fig. \ref{fig8}(d). The key feature remains the same as above. 
 
}

\section{Conclusions}

We have resolved the long-overlooked discrepancy in the time-dependent
Liouville equation between Kohn and Luttinger \cite{kohn1957} and
Aversa and Sipe \cite{aversa1995}. We have shown that the generalized
Liouville equation under LG has two partial derivatives with respect
to the crystal momentum and the density matrix is a full matrix
between two different $\bk$, fully consistent with Kohn and Luttinger
\cite{kohn1957}. The discrepancy emerges when the position operator is
assumed to be $\bk$-diagonal in CMR. The $\bk$-diagonal Liouville
equation misses all the off-diagonal elements of the density matrix
and may become incomplete when these elements are significant.  As
shown in our examples, the polarization can be both qualitatively and
quantitatively different. The $\bk$-diagonal formalism produces a
polarization that is 7 times smaller.  Thus our study establishes a
new paradigm for the time-dependent Liouville equation currently in
use, and will have a significant impact on both the past and future
research.  Our theory opens the door to ultrafast dynamics in real
materials at the first-principles level.

 \acknowledgments GPZ was partly supported by the U.S.\ Department of
 Energy under Contract No.\ DE-FG02-06ER46304. We acknowledge that our
 work was carried out on Indiana State University's Quantum and
 Obsidian Clusters. MSS was supported by the National Science
 Foundation of China under Grant No. 11874189.

$^*$guo-ping.zhang@outlook.com.
https://orcid.org/0000-0002-1792-2701

$^\dagger$ mingsusi@outlook.com


\appendix

\section{Commutator between the position and Hamiltonian}

We notice in their paper that AS invoked a well-known commutator
\cite{aversa1995}, \be [\br,\hh]=\frac{i\hbar}{m_e}\hp, \label{eq300}
\ee which is always valid as far as $\hh$ does not include a nonlocal
potential. But there is another hidden difficulty that receives much
less attention. The commutator, as it stands, does not allow to have a
one-to-one mapping from $\hp$ to $\br$.  Suppose we have an unknown
operator $\Lambda$ that commutes with $\hh$, i.e., $[\Lambda,\hh]=0$.
\be [\br-\Lambda,\hh]=\frac{i\hbar}{m_e}\hp.
   \label{eq301}
\ee If we compare Eqs. \ref{eq300} and \ref{eq301}, we see the right
side is exactly the same, but the left sides are different.  In CMR,
Eq. \ref{eq301} is \ba \la n\bk|\br|m\bk'\ra = \la
n\bk|\Lambda|m\bk'\ra +\frac{i\hbar}{m_e} \frac{\la n\bk|\hp|m
  \bk'\ra}{E_m(\bk')-E_n(\bk)} \breaker = \la n\bk|\Lambda|m\bk'\ra
+\frac{i\hbar}{m_e} \frac{\la n\bk|\hp|m \bk\ra \delta(\bk-\bk')}
{E_m(\bk')-E_n(\bk)} , \ea which shows that even if $\hp$ is diagonal
in $\bk$, it does not necessarily imply $\br$ is diagonal in $\bk$.
This shows that mapping from $\hp$ to $\br$ is a one-to-many mapping.

We can show another example. 
Using  CMR, Eq. \ref{eq300} becomes \be \la
n\bk|\br|m\bk'\ra (E_m(\bk')-E_n(\bk)) = \frac{i\hbar}{m_e} \la
n\bk|\hp|m \bk\ra \delta(\bk-\bk'). \label{eq300a} \ee Since $\hp$ is
translationally invariant, it is diagonal in $\bk$, $\la n\bk|\hp|m
\bk'\ra= \la n\bk|\hp|m \bk\ra \delta(\bk-\bk')$. Does this imply that
$\br$ is also $\bk$-diagonal?  This can be easily checked within a
simple one-dimensional free-electron model.  Consider two eigenstates,
$|k_1\ra=\frac{1}{\sqrt{L}}e^{ik_1x}$ and
$|k_2\ra=\frac{1}{\sqrt{L}}e^{ik_2x}$, which are orthonormalized
within a line segment of $x=[0,L]$. The exact matrix element of the
operator is defined through the integral $\la k_1|x|k_2\ra =\int_0^L
\frac{1}{L}e^{-ik_1x}xe^{ik_2x} dx.  $ Table \ref{table1} compares the
matrix elements computed directly by integration with those computed
through Eq. \ref{eq300a}.  It is clear that they do not agree with
each other. In particular, the position operator has nonzero
off-diagonal elements, in agreement with our finding above,
$\nabla_\bq\delta(\bq-\bk)\ne \delta (\bq-\bk)$, see the discussion
below  Eq. \ref{eq10}.

\begin{table}
  \caption{Comparison of the position matrix elements $\la
    k_1|x|k_2\ra$ 
    between the exact formalism and Eq. \ref{eq300a}.}
\begin{tabular}{lrr}
  \hline
    \hline
  ($k_1,k_2$) & exact & Eq. \ref{eq300a} \\
\hline
  $k_1= k_2$ & $\frac{L}{2}$ & undetermined  \\
$k_1\ne k_2$ & $\frac{1}{i(k_2-k_1)}$ & 0 \\
 \hline  \hline
\end{tabular}
\label{table1}
\end{table}

\end{document}